\documentclass[12pt]{iopart}
\usepackage{amsfonts}
\usepackage{amsthm}
\usepackage{graphics}
\usepackage{graphicx}
\usepackage{subfigure}
\usepackage{amssymb}
\usepackage{graphics}
\usepackage{graphicx}
\usepackage{epstopdf}
\usepackage{color}
\usepackage{subfigure}
\usepackage{pgfplots,tikz}
\usepackage{url}
\usepackage{float}
\usepackage{soul}
\def\x{{\bf x}}

\def\k{\textbf{k}}

\def\kmax{k_{\rm max}}
\newcommand\bet{{g}}
 
\newcommand\alps{{\frac{\hbar^2}{2m}}}

\newcommand\dertt[1]{ \frac{\partial{ #1}}{\partial t} }
\newcommand\gd{\mbox{${\bf \nabla}^{2}$}}
\newcommand\grad{\mbox{${\bf \nabla}$}}
\newcommand\psib{\overline{\psi}}

\providecommand{\keywords}[1]{{\textit{Keywords: }} #1}

\def\TBKT{T_{\rm BKT}}
\def\Np{{N_{\rm I}}}
\def\Vp{V_{\rm I}}
\def\Mp{M_{\rm I}}
\def\Rp{a_{\rm I}}
\def\Epp{E_{\rm I \leftrightarrow I}}
\def\rmin{r_{\rm min}}

\begin{document}

\title{Clustering and phase transitions in a 2D superfluid with immiscible active impurities}
\author{Umberto Giuriato$^1$, Giorgio Krstulovic$^1$ and Davide Proment$^2$ }
\address{$^1$ Universit\'e C\^ote d'Azur, Observatoire de la C\^ote d'Azur, CNRS, Laboratoire Lagrange, Bd de l'Observatoire, CS 34229, 06304 Nice cedex 4, France.}
\address{$^2$ School of Mathematics, University of East Anglia, Norwich Research Park, Norwich, NR4 7TJ, United Kingdom}

\begin{abstract}
Phase transitions of a finite-size two-dimensional superfluid of bosons in presence of active impurities are studied by using the projected Gross--Pitaevskii model. Impurities are described with classical degrees of freedom. A spontaneous clustering of impurities during the thermalization is observed. Depending on the interaction among impurities, such clusters can break due to thermal fluctuations at temperatures where the condensed fraction is still significant. The emergence of clusters is found to increase the condensation transition temperature. The condensation and the Berezinskii--Kosterlitz--Thouless transition temperatures, determined numerically, are found to strongly depend on the volume occupied by the impurities: a relative increase up to a $20\%$ of their respective values is observed, whereas their ratio remains approximately constant.
\end{abstract}


\maketitle

\section{Introduction}

When a fluid composed of bosons is cooled down or the number of particles is increased, the system experiences a phase transition giving rise to a macroscopic state known as Bose--Einstein condensate (BEC) \cite{griffin1996bose}.
Since its first experimental observation by Anderson et al. \cite{BECExp95}, BECs have been realized in systems of very different nature such as cold atomic gases \cite{griffin1996bose, Hadzibabic}, solid-state quasiparticles \cite{kasprzak2006bose, demokritov2006bose} and light in optical micro-cavities \cite{klaers2010bose}. 
Whereas in three spatial dimensions a condensate is stable with respect to thermal fluctuations, in two dimensions such fluctuations can destroy the long-range order of the system. 
This is a general result in statistical field theory known as the Mermin--Wagner--Hohenberg theorem \cite{MerminWagner,Hohenberg}: it states that a continuous symmetry cannot be spontaneously broken in dimensions lower than three, otherwise large-scale Goldstone modes would have an infinite infrared contribution to the two-point correlator. 
This theorem assumes the thermodynamic limit, that is the system size being infinite. 
However, for a finite system, condensation can be recovered, having a transition temperature $T_\lambda$ that vanishes as the inverse of the logarithm of the system size.

Although condensation is formally forbidden in an infinite two-dimensional system, a peculiar phase transition of a different nature has attracted the attention of physicists and mathematicians since its independent discovery in the early 70's by Berezinskii,  Kosterlitz and Thouless (BKT) \cite{Kosterlitz1,Kosterlitz2,berezinskii1971destruction}. 
The BKT transition is an infinite-order topological phase transition and manifests itself in systems that belong to the same universality class. 
By approaching the BKT transition temperature, $\TBKT$, from below, the system switches from a gas of bounded vortex-antivortex pairs to a gas of free vortices, moving from a quasi-ordered phase to a disordered one. 
The BKT transition has been observed in BECs made of dilute gases \cite{Experiment1, Experiment2, Experiment3, Experiment4, Experiment5}, exciton-polaritons \cite{ExcitonPolaritons1}, liquid helium films \cite{BishopHelium1} and studied theoretically and numerically \cite{Svistunov, BKTBECHarmonic, DavideBKT,shukla2013turbulence}; for a review on the topic, see for instance \cite{Hadzibabic}.

The purpose of this article is to study how the statistical mechanical properties of a two dimensional superfluid of bosons are affected by the presence of impurities.
Particles and impurities have been used in superfluids since the early experiments in $^4$He \cite{donnelly1991quantized} mainly with detection purposes: electrons, ions and neutral impurities such as hydrogen particles and excimers have been exploited to visualize quantized vortices, to study their dynamics and the statistics of superfluid (quantum) turbulence \cite{bewley2006superfluid, ZmeevTracersInHe4, LaMantiaParticles}.
More recently, the investigation of the interaction between one or more impurities and superfluids has been the main topic of experiments in cold atoms \cite{ParticleInBECSpethman, 1DBECImpuritiesCatani}, superfluids of light \cite{ParticleFluidoffLight,CarusottoLightParticles} and polaritons in semiconductor microcavities \cite{AmoPolaritons}.
 On the theoretical side, the dynamics of impurities in a BEC has been also addressed  \cite{QuantumCluster}, as well as the properties of $^3\rm He$ and $\rm H$ impurities on thin $^4\rm He$ films \cite{HeliumTheory1, HeliumTheory2}. In addition, Rica \& Roberts studied how a collection of impurities affects the ground-state of a BEC by using a mean field model \cite{RicaRoberts}. In this last work, four phases were identified, depending on the value of the interaction couplings. In particular, if the scattering lengths between impurity fields are positive, impurities behave as localized objects, they separate from the condensate and present a hard-sphere repulsion between each other.

We investigate here impurity clustering and phase transitions occurring in a minimal model that mimics such situation: the Gross--Pitaevskii (GP) equation coupled with active immiscible impurities having classical degrees of freedom. 
Such model was introduced in \cite{ActiveWiniecki} and recently used in two-dimensional numerical simulations to study impurity-impurity and impurity-vortex interactions \cite{ShuklaParticlesPRA2016,ShuklaParticlesPRA2017}. 
Finite-temperature BECs can be studied by using the projected GP equation, that is obtained by introducing a cut-off $k_{\max}$ in Fourier space: this regularizes the classical mean-field ultra-violet divergence. 
The projected GP model is an effective model to study the condensation transition in two and three dimensions \cite{DavisFiniteTEmpBEC,CondensationRica,KrstulovicBottleneck,DavideBKT} and superfluid vortex dynamics at finite temperature \cite{BerloffRing,KrstulovicSlowdown,KrstulovicLongPREFiniteTemp}. 

\section{Theoretical model and numerical results}

\subsection{Model for impurities in a superfluid}
We generalize the projected GP model to include the dynamics of active impurities. 
The model is then described by the Hamiltonian
\begin{eqnarray}
\nonumber H&=&\int\left( \frac{\hbar^2}{2m} |\grad \psi |^2 +\frac{g}{2}|\mathcal{P}_{\rm G}[|\psi|^2]|^2 +\sum_{i=1}^{\Np} \Vp(| \x -{\bf q}_i |)\mathcal{P}_{\rm G}[|\psi|^2]\right) \mathrm{d} \x \\
 &&\hspace{5.8cm}+\sum_{i=1}^{\Np}\frac{{\bf p}_i^2}{2 \Mp}+ \sum_{i<j}^{\Np}V_{\rm rep}(|{\bf q}_i-{\bf q}_j|),
\label{Eq:HGP}
\end{eqnarray}
where $\psi$ is the collective wave-function of bosons having mass $ m $, and $g=4 \pi  a_\mathrm{s} \hbar^2 /m $ being $a_{\rm s}$ the $s$-wave scattering length of bosons interaction. $\Np$ is the total number of impurities of mass $ \Mp$, that are described using their classical position and momentum ${\bf q}_i$ and ${\bf p}_i$, respectively. 
The strong repulsive potential $\Vp$ determines the shape of the impurities by creating a large depletion in the fluid density. $V_{\rm rep}$ is a repulsive potential between impurities. 
The Galerkin projector $\mathcal{P}_{\rm G}$ truncates the system acting in Fourier space as $\mathcal{P}_{\rm G} \hat{\psi}_{\bf k}=\theta(\kmax-k)\hat{\psi}_{\bf k}$ with $\theta(\cdot)$ the Heaviside function, $\hat{\psi}_\k$ the Fourier transform of $\psi(\x)$ and $\k$ the wave vector.
The equations of motion are directly obtained by varying (\ref{Eq:HGP}):
\begin{equation}
i\hbar\dertt{\psi}=\mathcal{P}_{\rm G} [- \alps \gd \psi + \bet\mathcal{P}_{\rm G} [|\psi|^2]\psi+ \sum_{i=1}^{\Np}  \Vp(| \x -{\bf q}_i |)\psi]\label{Eq:GPE}
\end{equation}
\begin{equation}
\Mp\ddot{\bf q_i}= - \int  \Vp(| \x -{\bf q}_i |)  \mathcal{P}_{\rm G}[\nabla|\psi|^2]\, \mathrm{d} \x-\sum_{j\neq i}^{\Np}\nabla V_{\rm rep}(q_{ij}),\label{Eq:Particles}
\end{equation}
where  $q_{ij}=|{\bf q}_i-{\bf q}_j|$ and we have replaced ${\bf p}_i=\Mp\dot{\bf q}_i$. 
The previous set of equations exactly conserves the Hamiltonian, the number of bosons $N=\int |\psi|^2\mathrm{d} \x$ and momentum ${\bf P}=\int \frac{i\hbar}{2}\left( \psi {\bf \nabla}\psib - \psib {\bf \nabla}\psi\right)\mathrm{d} \x+\sum_i {\bf p}_i$.

The impurities in the system feel an attractive force mediated by the superfluid density field \cite{RicaRoberts,ShuklaParticlesPRA2016}. 
However, unlike the case of impurities described by classical fields \cite{RicaRoberts}, in the model (\ref{Eq:HGP}) no repulsion mediated by the fluid exists. In order to mimic such a hard-sphere repulsion, we consider the Lennard--Jones-like potential $ V_{\rm rep}(r)=\epsilon \rmin^{12}/r^{12}$,  as in \cite{ShuklaParticlesPRA2016}. We fix the energy $\epsilon$ in order to set the minimum of the total interaction energy between impurities at zero temperature at a distance $r_\mathrm{min}$. Note that the specific shape of $V_\mathrm{rep}$ is not important, as long as it reproduces a hard-sphere repulsion.
For the impurities potential we use a smoothed hat-potential $\Vp(r)=V_0(1-\tanh\left[(r^2 -\Rp^2)/4\Delta l^2\right])/2$, where $\Rp$ sets the characteristic radius of the impurity and $\Delta l$ is a smoothing parameter. 
Finally, let us notice that in absence of impurities and at zero temperature, eq.(\ref{Eq:GPE}) can be linearized about a uniform density state $ |\psi|^2=\rho_\infty/m $, defining the phonon (sound) velocity $c=\sqrt{g\rho_\infty/m^2}$ with dispersive effects taking place for length scales smaller than the healing length $\xi=\sqrt{\hbar^2/2g\rho_\infty }$.

We integrate the system (\ref{Eq:GPE}-\ref{Eq:Particles}) by using a pseudo-spectral code with $N_{res}$ uniform grid points per direction of a squared domain of size $L=2\pi$.
We set $\kmax=N_{res}/3$ so that the truncated system exactly conserves all the invariants (provided that initially $\mathcal{P}_{\rm G} \psi=\psi$ and $\mathcal{P}_{\rm G} \Vp=\Vp$) \cite{KrstulovicLongPREFiniteTemp},  $c=\rho_\infty=1$, $V_0=10$ and $\epsilon =0.00674$. 
As the healing length changes with temperature, we parametrize the solutions of (\ref{Eq:GPE}-\ref{Eq:Particles}) using its value taken at zero temperature. 
In thermal states, the only relevant dimensionless parameters are $L/\xi$, $\Rp/\xi$, $\Np$, the relative mass $\mathcal{M}=\Mp/\rho_\infty\pi\Rp^2$ and $\xi\kmax$. We set $L/\xi=128$.
The value of $\xi\kmax$ controls the strength of the nonlinear interactions and it is kept fixed to $\xi\kmax=2\pi/3$. 
For this value, most of the excitations are phonons when the condensate fraction is large. 
For instance, it is compatible with the one used in \cite{DavisFiniteTEmpBEC}, that is $\xi\kmax\sim2$; such value applies to a gas of $^{87}$Rb atoms.

\begin{figure*}
\centering
\includegraphics[width=1\linewidth]{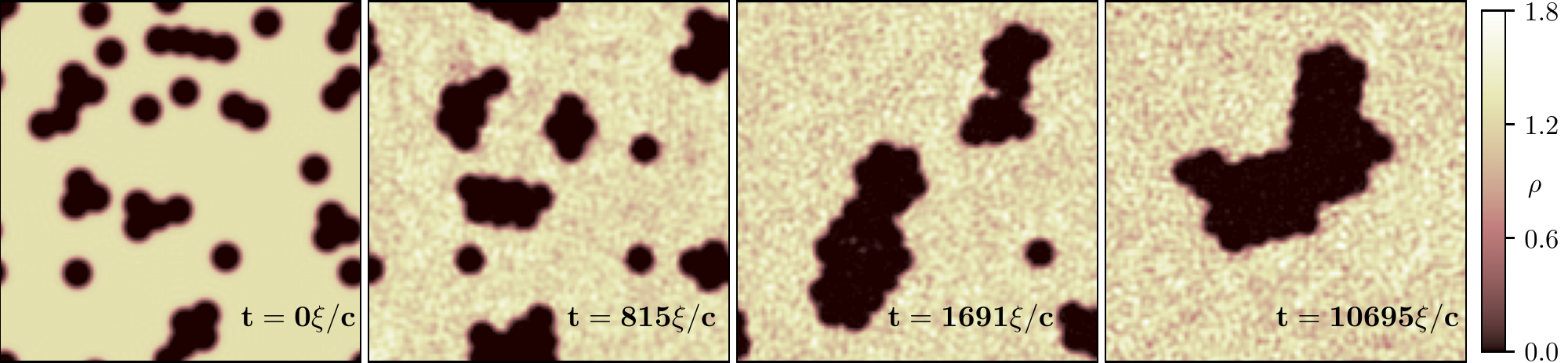}
\caption{(\textit{Color online}) Snapshots of the fluid density during the GP temporal evolution of a state with 40 impurities (dark holes).}
\label{Fig:ClusteringThermal}
\end{figure*}

We start by presenting a long temporal evolution of a system having 40 impurities of mass $\mathcal{M}=0.1$, radius $\Rp=4\xi$, initially located at random positions (avoiding overlaps) and having zero velocity.
The density field of the initial condition is displayed in figure \ref{Fig:ClusteringThermal} ($t= 0 \xi/c$).
Impurities correspond to dark holes.
During the time evolution, the short-range interaction among impurities mediated by the fluid let them collapse into small clusters ($t= 815 \xi/c$); waves with random phases are generated, populating small length scales and starting the thermalization process.
This thermal noise induces the clusters to move in a stochastic way, and to grow further ($t= 1691 \xi/c$). 
Eventually, the system reaches thermal equilibrium where only one big cluster is observed in a bath of thermalized waves, ($t= 10965 \xi/c$). A movie of the the evolution is available in the Supplementary Data.

\subsection{Grand-canonical thermal states}

The evolution illustrated in figure \ref{Fig:ClusteringThermal} is an example of thermalization occurring in the micro-canonical ensemble, as the thermal state is achieved keeping all the invariants conserved. 
Such dynamical process is numerically costly and does not directly provide access to the conjugate thermodynamical variables: temperature and chemical potential (here we only consider zero momentum states).
To overcome these issues, in \cite{KrstulovicLongPREFiniteTemp} a stochastic relaxation was introduced in order to efficiently generate thermal states in the grand-canonical ensemble. 
We make use of this approach adapting it to the Hamiltonian (\ref{Eq:HGP}).  
The stochastic dynamics is ruled by
\begin{equation}
\hbar\frac{\partial\hat{\psi}_\mathbf{k}}{\partial t}=-\frac{\partial F}{\partial \hat{\psi}_\mathbf{k}^*} + \sqrt{\frac{2\hbar}{\beta}}\hat{\xi}_\mathbf{k}\label{eq:stocpsi}\\
\end{equation}
\begin{equation}
\frac{\partial {\bf q}_{i}}{\partial t}=-\frac{\partial F}{\partial  {\bf q}_{i}} + \sqrt{\frac{2}{\beta}}{\bf \xi}_i^q,\quad\Mp\frac{\partial \dot{{\bf q}}_i}{\partial t}=-\frac{\partial F}{\partial\dot{{\bf q}}_i} + \sqrt{\frac{2\Mp}{\beta}}{\bf \xi}_i^{\dot{q}}\label{eq:stocq}
\end{equation}
where $F=H-\mu N$ is the free energy of the system,  $\mu$ is the chemical potential controlling the number of bosons and $\beta $ is the inverse temperature; $\hat{\xi}_\mathbf{k},{\bf \xi}_i^q$ and ${\bf \xi}_i^{\dot{q}}$ are independent Gaussian white noises of unit variance. 
It can be shown by using the Fokker-Planck equation associated to (\ref{eq:stocpsi}-\ref{eq:stocq}), that the stationary probability distribution is given by the Gibbs grand-canonical distribution $\mathbb{P}[\hat{\psi}_\mathbf{k},{\bf q}_i,{\bf \dot{q}}_i]\propto e^{-\beta F}$. 
In the micro-canonical ensemble, $\mathbb{P}[\hat{\psi}_\mathbf{k},{\bf q}_i,{\bf \dot{q}}_i]$ is also the stationary solution of the Liouville equation that describes the evolution of the phase-space distribution of the Hamiltonian system (\ref{Eq:HGP}) \cite{KrstulovicLongPREFiniteTemp}. It is evident from (\ref{eq:stocq}) that varying the impurity masses modifies only the variance of impurity momenta in the steady state. Namely, the configurations of impurities in the steady state and the statistical properties of the thermalized system are independent of impurity masses.
We define the temperature as $T=1/k_\mathcal{N}\beta$, with $k_\mathcal{N}=L^2/\mathcal{N}$ and $\mathcal{N}=\pi \kmax^2$ the total number of Fourier modes. With this definition $T$ is an energy per unit of surface such that at low temperatures $F\approx T L^2$, because of equipartition. 
With these choices, intensive quantities remain constant when increasing the system size. 
In addition, we fix the total density mass $\bar{\rho}=mN/L^2=1$ by dynamically adjusting the chemical potential \cite{KrstulovicLongPREFiniteTemp}.
We use (\ref{eq:stocpsi}-\ref{eq:stocq}) to study the effect of impurities on the superfluid condensed fraction $n_0$ defined as
\begin{equation}
n_0=\frac{\langle\left| \int_{\mathcal{V}'} \psi({\bf x})\,\mathrm{d}{\bf x}\right|^2 \rangle_T}{ \langle\int_{\mathcal{V}'}| \psi({\bf x})|^2\,\mathrm{d}{\bf x}\rangle_{T}} \label{Eq:CondFrac},
\end{equation}
where $\mathcal{V}'$ is the domain excluding the region occupied by the impurities and $\langle \cdot\rangle $ stands for average over realizations at temperature $T$
\footnote{This definition gives the same result as 
$
n_0= \left(\frac{\left|\langle \psi\rangle_T\right|^2}{N}\right)\left(\frac{\left|\langle \psi\rangle_{T=0}\right|^2}{N}\right)^{-1} = \frac{\left|\langle \psi\rangle_T\right|^2}{\left|\langle \psi\rangle_{T=0}\right|^2}
$
. For numerical convenience we use the latter in our computations.}.

\subsection{Clustering of impurities}
\begin{figure}
\centering
\includegraphics[width=1.\linewidth]{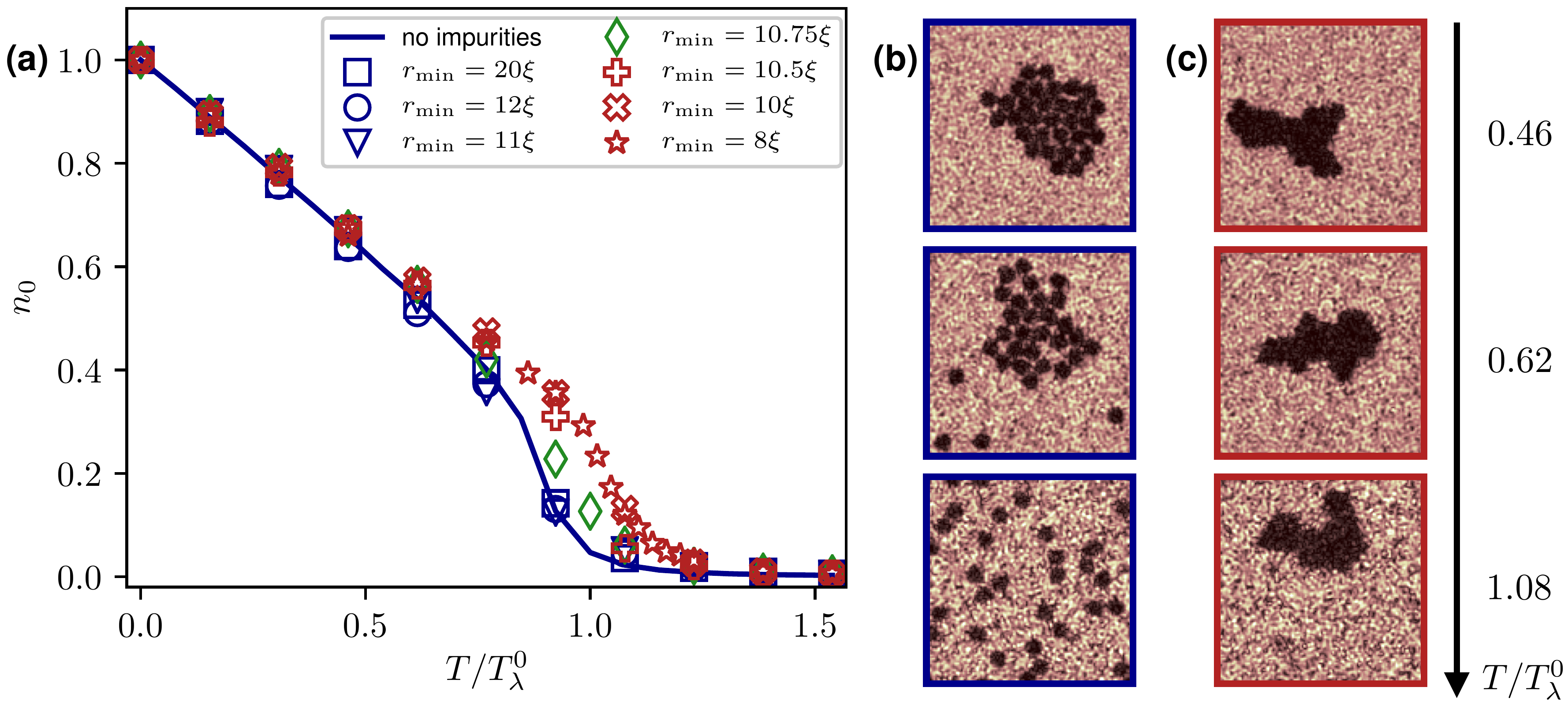}
\caption{(\textit{Color online})
 \textbf{(a)}: Condensed fraction as a function of the temperature for different values of $r_{\rm min}$. Temperatures are expressed in units of the condensation temperature with no impurities $T_\lambda^0$.  \textbf{(b)}: Snapshots of the density field in the steady state at different temperatures in the case of high repulsion ($r_\mathrm{min} = 11\xi$) among impurities.  \textbf{(c)}: The same of \textbf{(b)} but for low repulsion ($r_\mathrm{min} = 8\xi$).
Scans are performed with $\Np=31$ and $\Rp=4\xi$. }  
\label{Fig:ClusteringPhase}
\end{figure}

We first perform a temperature scan without impurities. The condensed fraction is shown in figure \ref{Fig:ClusteringPhase}(a) (solid blue line). The transition temperature $T_\lambda$ is the lowest temperature where the condensed fraction can be considered negligible. We estimate it in a consistent way adopting the following numerical protocol: we take the points around which $n_0^2(T)$ is close to zero and we perform a linear interpolation of it. $T_\lambda$ is then determined by finding the point where the linear fit vanishes. From now on, we indicate with $T_\lambda^0$ the transition temperature in the system without impurities.
Then, we perform temperature scans varying the repulsive potential parameter $\rmin$ with a fixed number of impurities $\Np=31$ having radius $\Rp=4\xi$.
The results are also shown in figure \ref{Fig:ClusteringPhase}. In figures \ref{Fig:ClusteringPhase}(b) and \ref{Fig:ClusteringPhase}(c) snapshots of  in the steady date are displayed respectively in the case of high and low repulsion among impurities. For both cases we report three different temperatures. Depending on the strength of the repulsion potential two different behaviors of $n_0$ can be observed, as it is clear in figure \ref{Fig:ClusteringPhase}(a). When the repulsion among impurities is strong enough (blue markers, $\rmin\ge11\xi$), clusters are broken already at temperatures lower than $T_\lambda^0$ (see figure \ref{Fig:ClusteringPhase}(b)) and impurities have no appreciable effect on $n_0$. On the other hand, for $\rmin\le10.5\xi$ (red markers) impurities remain clustered  and lead to an increase of the  $n_0$ at medium-high temperatures (see figure \ref{Fig:ClusteringPhase}(c)). 

\begin{figure}
\centering
\includegraphics[width=1.\linewidth]{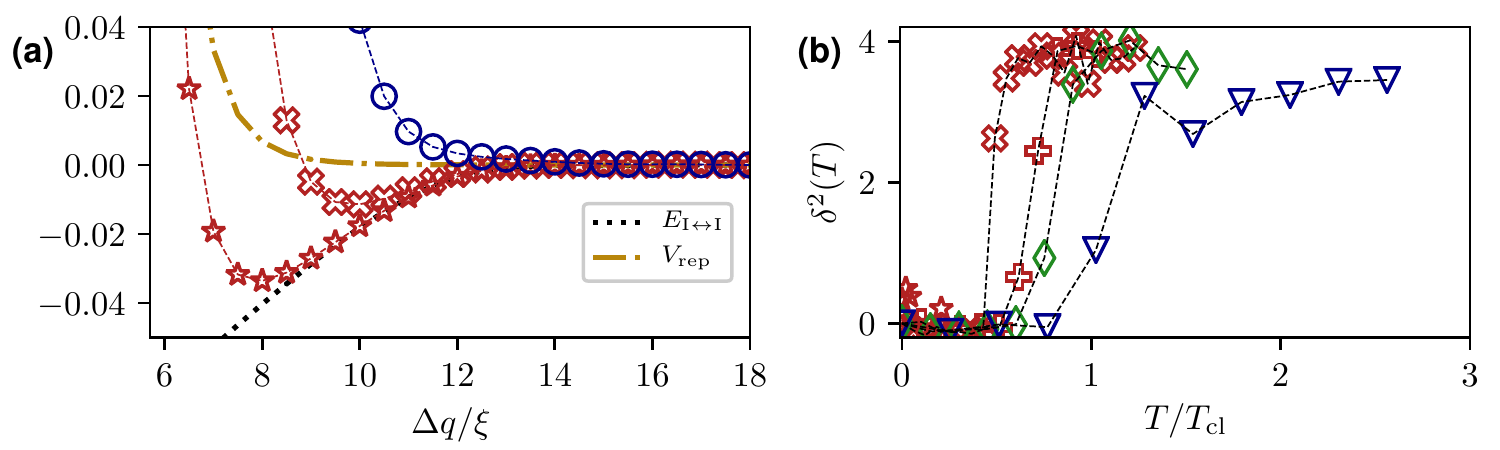}
\caption{(\textit{Color online})
 \textbf{(a)}: Impurity-impurity interaction $\Epp$ as a function of their distance $\Delta q/\xi$ (dotted black line) and repulsive potential  (dotted-dashed golden line) for $\rmin=8\xi$. Different markers correspond to total energy $\Epp+ V_{\rm rep}$ for different values of $\rmin$.
 \textbf{(b)}: Relative impurity distance $\delta^2(T)$ as a function of the normalized temperature  $T/T_{\rm cl}$.
Scans are performed with $\Np=31$ and $\Rp=4\xi$. The ratios $T_\mathrm{cl}/T_\lambda$ are $7.41,\,2.51,\,1.5,\,1.02,\,0.53,\,0.002$ for $r_\mathrm{min}$ from $8\xi$ to $12\xi$ respectively.  The markers refer to the same legend as in figure 2.}  
\label{Fig:ClusteringTrans}
\end{figure}

It has been shown that impurities experience a short-range attractive force, mediated by the superfluid density \cite{RicaRoberts,ShuklaParticlesPRA2016}. This interaction is characterized by a potential energy, denoted here  $\Epp$. We compute this energy numerically by measuring the full GP free energy of the ground state with two impurities placed at a distance $\Delta q$, without the contribution of the repulsion. The constant contribution to $\Epp$ is eventually set to zero. 
Impurities are then repelled because of $V_{\rm rep}$, generating cluster structures as the one observed in figure \ref{Fig:ClusteringThermal}(d). 
However, if thermal fluctuations are large enough, the bound among impurities can be broken. In figure \ref{Fig:ClusteringTrans}(a) we compute the interaction energy between two impurities at zero temperature $\Epp$, as a function of their distance $\Delta q$ (dotted black line). 
As a reference, the figure also displays the repulsive potential $ V_{\rm rep}$ with $\rmin=2\Rp$ (dotted-dashed golden line). The sum of both potentials is displayed in the same figure for different values of $\rmin$: for sufficiently small values of $ \rmin $, a potential well $\Delta U$ centered at $\rmin$ appears. 
We thus expect a pair of impurities to split in a finite time at the temperature $ T_{\rm cl} \sim \Delta U / k_\mathcal{N} $, as in a standard escape problem from a potential well \cite{gardiner2009stochastic}.
In order to quantify this clustering transition, we measure the average square distance between impurities and their center of mass
\begin{equation}
\delta^2(T)=\frac{d^2(T)-d^2(0)}{d^2(0)},\quad{\rm with }\quad d^2(T)=\sum_{j=1}^\Np\langle |{\bf q}_j-{\bf q}_\mathrm{cm}|^2\rangle \quad{\rm and }\quad \mathbf{q}_\mathrm{cm}=\sum_j^{N_I}\frac{\mathbf{q}_j}{N_\mathrm{I}}.
\label{Eq:delta2}
\end{equation}
Figure \ref{Fig:ClusteringTrans}(b) displays $\delta^2(T)$ as a function of $T/T_{\rm cl}$ for different values of $\rmin$.
A transition around $T_{\rm cl}$ is indeed observed, where discrepancies are likely due to oversimplifications made in the estimation of $T_\mathrm{cl}$, namely by neglecting the many-body impurity interactions and by using the interaction potential obtained at $T=0$. 
For weak repulsion, even if the condensate vanishes, impurities still feel the density-mediated attraction.

\subsection{Condensation and BKT transition temperatures in presence of impurity clusters}

\begin{figure}
\centering
\includegraphics[width=1.\linewidth]{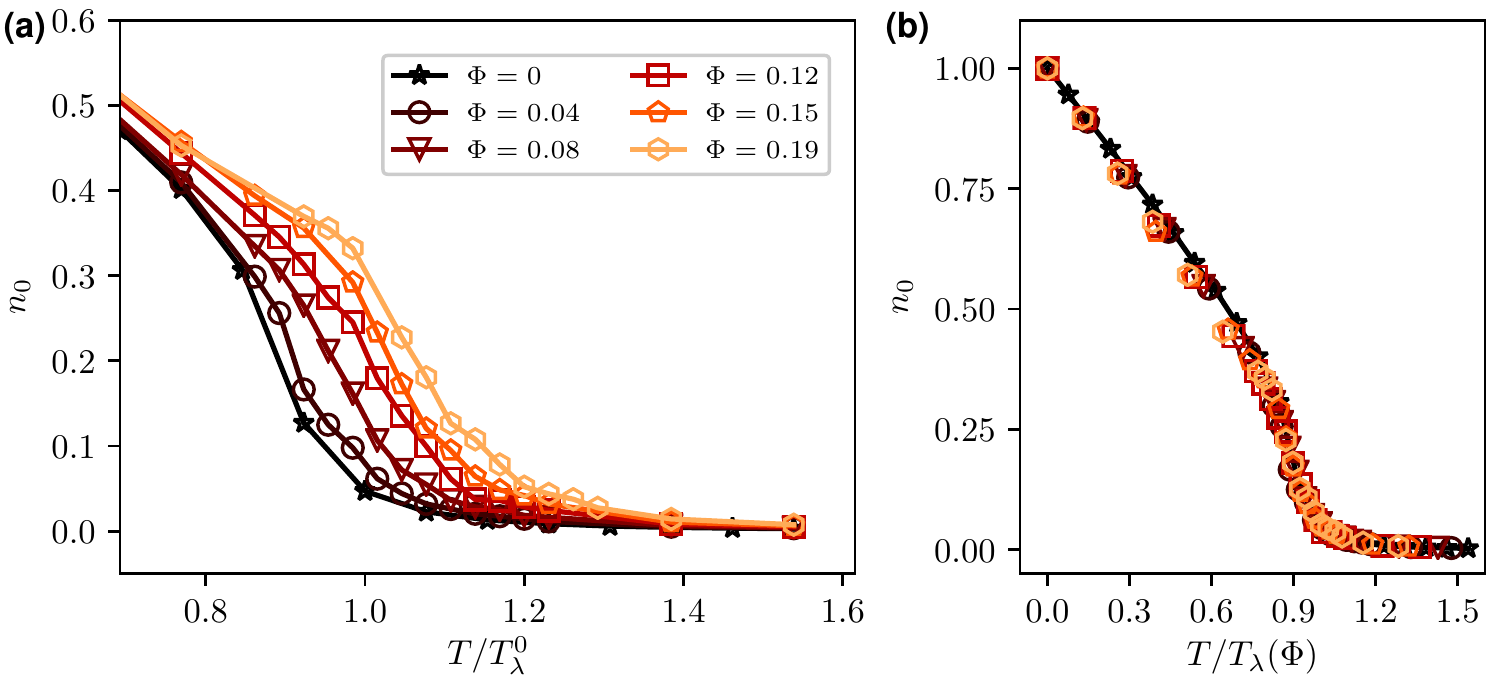}
\caption{(\textit{Color online}) \textbf{(a)}: Condensed fraction as a function of temperature for different values of the filling fraction $\Phi$. Temperatures are expressed in units of the condensation temperature with no impurities $T_\lambda^0$.
 \textbf{(b)}: Condensed fraction as a function of the temperature normalized with $T_\lambda(\Phi)$.  Scans with $r_p=4\xi$ and $r_\mathrm{min}=2r_\mathrm{p}=8\xi$.}
\label{Fig:Condensation}
\end{figure}

Studying different values of $r_\mathrm{min}$ allowed us to show that an increasing of $n_0$ occurs only when the fluid depletion, due to the presence of impurities, is confined to a large connected region at all temperatures. Therefore such effect can not be simply explained by the local increase of density in regions not occupied by impurities. In the following we consider hard-sphere impurities by fixing $r_\mathrm{min}=2\Rp$.
In order to quantitatively characterize the change in $n_0$, we study how the condensation transition changes when varying the filling fraction 
\begin{equation}
\Phi=1 - \frac{ |\langle \psi \rangle_{T=0}|^2} {\langle|\psi|^2\rangle_{T=0}}, 
\label{Eq:fillingfraction}
\end{equation}
which corresponds to the fraction of the total volume occupied by the impurities. In figure \ref{Fig:Condensation}(a) the condensed fraction is shown for different values of $\Phi$, obtained by varying the number of impurities. 
It is evident that the larger is the number of clustered impurities, the higher results the condensation transition temperature. We explicit the dependence of the transition temperature on the filling fraction as $T_\lambda(\Phi)$.

\begin{figure}
\centering
\includegraphics[width=1.0\linewidth]{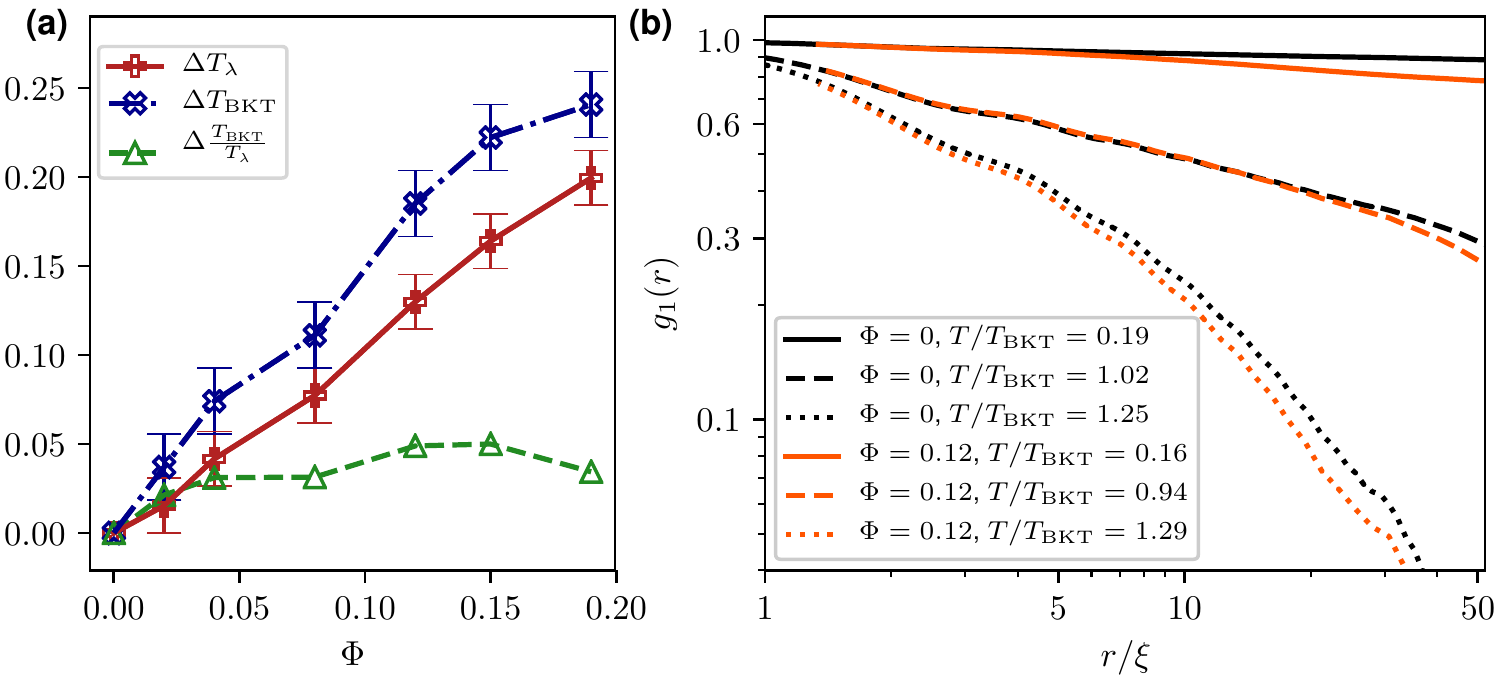}
\caption{(\textit{Color online}) \textbf{(a)}: Relative increments of $T_\lambda$ and $T_{\rm BKT}$ as a function of the filling fraction $\Phi$. The relative increment of the ratio $\frac{T_\mathrm{BKT}}{T_\lambda}$ is also shown. 
\textbf{(b)}: Spatial correlation function $g_1(r)$ for three different temperatures (lower, close and higher than $T_\mathrm{BKT}$) with and without impurities.}
\label{Fig:Temperatures}
\end{figure}

The condensation temperature $T_\lambda(\Phi)$ is measured for different values of $\Phi$ following the same procedure explained in the previous section. In figure \ref{Fig:Temperatures}(a) the relative increase $\Delta T_\lambda=(T_\lambda(\Phi)-T_\lambda^0)/T_\lambda^0$ is displayed. 
Remarkably, $\Delta T_\lambda$ scales linearly with $\Phi$ growing up to $20\%$. 
We have checked by varying the number of impurities, their size and the parameter $r_\mathrm{min}$ for values lower than $2a_\mathrm{I}$, that $n_0$ only depends on $\Phi$ and $T$ (data not shown).
Despite the change on $T_\lambda$, the condensed fraction curves collapse as expected to a single one, once plotted versus $T/T_\lambda(\Phi)$ (see figure \ref{Fig:Condensation}(b)). 

Finally, we briefly address the role of impurities in the the BKT transition. 
A detailed study will be left for a further work. 
This phase transition manifests through a change in the behavior of the correlation function $g_1(r)=\langle \psi(0)\psi^*({\bf r})\rangle$ at the BKT transition temperature $T_{\rm BKT}$. 
At $T<T_{\rm BKT}$, it presents a power-law decay $g_1(r)\sim r^{-\alpha}$, where $\alpha$ depends linearly on the temperature; at high temperatures, it exhibits the standard exponential decay of disordered systems. 
In figure \ref{Fig:Temperatures}(b) we show $g_1(r)$ at different temperatures (lower, close and higher than $T_\mathrm{BKT}$) where these two behaviors are clearly distinguishable. The BKT transition temperature can be thus determined by finding where $g_1(r)$ abruptly changes its behavior \cite{AustralianBKT}. 
With no impurities  and the parameters used in this article, the BKT transition takes place at $T_{\rm BKT}=0.83T_\lambda^0$.
Note that because of the Mermin--Wagner--Hohenberg theorem \cite{MerminWagner,Hohenberg}, the condensation critical temperature vanishes as $1/\log{L}$ in the thermodynamic limit, so in principle for a very large system we could have $T_\lambda^0<T_{\rm BKT}$. We do not address such limit in this article. 
The presence of impurities in the system modifies the decay of $g_1(r)$ by shifting $T_{\rm BKT}$ to higher temperatures.
Figure \ref{Fig:Temperatures}(a) also displays the relative increase $\Delta T_{\rm BKT}=(T_{\rm BKT}(\Phi)-T_{\rm BKT}^0)/T_{\rm BKT}^0$ of the BKT transition temperature for different filling fractions $\Phi$. Although $T_\lambda$ and $T_{\rm BKT}$ both grow up to $20\%$ when the $\Phi$ is increased, their ratio remains almost constant. 
Let us remark that there are no important effects on $T_\mathrm{BKT}$ if impurities are not clustered.

The increase of the transition temperatures $T_\lambda$ and $T_{\rm BKT}$ can be explained by a simple phenomenological argument. 
Large objects in the system, such as the clustered impurities, modify the fluid wave-function boundary conditions. In particular, they impose effective Dirichlet boundary condition leading to symmetries in Fourier space, decreasing the number of active modes. At a given temperature, with less active modes, the energy is smaller and thus a higher temperature is necessary to induce a transition.  We stress that this result is general and does not depend on the choice of the repulsion potential $V_\mathrm{rep}$, as long as it is sufficiently short-range to allow the formation of a large-size cluster at all $T<T_\lambda$. Our results could apply as well to two-component BECs, with the components having different condensation temperatures and strong repulsion between them. Finally, the same effect on the condensation curve will occur in three dimensions, as it comes from a geometrical effect.

\section{Discussion}

In this article we studied thermal states of two-dimensional superfluids with active impurities.
We demonstrated how the phase transitions are affected by the emergence of impurity clusters, opening up the possibility to raise the transition temperatures in experiments by doping superfluids with specific types of impurities. Such result rises new questions that would be interesting to address in detail. In particular, it is remarkable that the presence of impurities does not disorder the system inducing a loss of coherence. Is there a maximum value of the critical temperature that can be reached using impurities, or it will continue to increase until the impurities occupy the full domain? Could the modification of the condensation curve be rephrased as a competition between the full perimeter and the full area of the impurities?
Moreover, this system presents a rich behaviour that, up our knowledge, has not yet been addressed in details. 
For instance, during the thermalisation dynamics, impurities cluster similarly to a diffusion-limited aggregation process \cite{DLAWitten}. 
Also, a complete study of the BKT transition, considering the opposite limit $T_\lambda<T_{\rm BKT}$, needs to be investigated and might devise new interesting physics. 

\begin{ack}  
The Authors were supported by the cost-share Royal Society International Exchanges Scheme (ref. IE150527) in conjunction with CNRS. 
GK and DP acknowledge the Federation Doeblin for supporting DP during his sojourn in Nice.
DP was supported by the UK Engineering and Physical Sciences Research Council (EPSRC) research grant EP/P023770/1. 
GK and UG were also supported by the ANR JCJC GIANTE ANR-18-CE30-0020-01.
Computations were carried out on the M\'esocentre SIGAMM hosted at the Observatoire de la C\^ote d'Azur. The authors acknowledge the anonymous referees for their interesting comments.
\end{ack}

\section*{References}

\end{document}